\documentclass[logo,a4paper,11pt]{ETHpaper}

\usepackage[utf8]{inputenc}

\usepackage{amsmath}
\usepackage{tikz}
\usetikzlibrary{calc}
\usetikzlibrary{decorations.markings}
\usetikzlibrary{arrows, arrows.meta}
\usetikzlibrary{positioning}
\usepackage[capitalise]{cleveref}

\usepackage{pgfplots}
\pgfplotsset{compat=newest}
\usepgfplotslibrary{groupplots}
\usepackage{nicefrac}
\usepackage{url}

\usepackage{enumitem}

\definecolor{colorSG}{RGB}{168,50,45}
\definecolor{customY}{HTML}{FBB13C}
\definecolor{customG}{HTML}{218380}
\definecolor{customT}{HTML}{73D2DE}
\definecolor{customW}{HTML}{FCFCFF}
\definecolor{customB}{HTML}{2E5EAA}
\definecolor{customP}{HTML}{5B4E77}
\definecolor{gitRedFaded}{HTML}{ffeef0}
\definecolor{gitGreenFaded}{HTML}{e6ffed}
\definecolor{gitRed}{HTML}{ffdce0}
\definecolor{gitGreen}{HTML}{cdffd8}
\definecolor{gitRedFull}{HTML}{cb2431}
\definecolor{gitGreenFull}{HTML}{2cbe4e}

\newlength{\w}
\newlength{\h}

\newcommand{\toolname}{\texttt{gambit}}
\newcommand{\zenodolink}{\url{http://doi.org/10.5281/zenodo.4384646}}

\title{gambit~-- An Open Source Name Disambiguation Tool for Version Control Systems}

\author{Christoph Gote$^{*}$ \qquad Christian Zingg$^{\dagger}$}
\authoralternative{C. Gote, C. Zingg}
\address{Chair of Systems Design, \\ ETH Zurich,
Weinbergstrasse 56/58, 8092 Zurich, Switzerland \\[2mm]
\(^*\)cgote@ethz.ch\\
\(^\dagger\)czingg@ethz.ch
}
\www{\url{http://www.sg.ethz.ch}}
\reference{Version of: \today}

\begin{document}

\maketitle

\begin{abstract}
Name disambiguation is a complex but highly relevant challenge whenever analysing real-world user data, such as data from version control systems.
We propose \toolname, a rule-based disambiguation tool that only relies on name and email information.
We evaluate its performance against two commonly used algorithms with similar characteristics on manually disambiguated ground-truth data from the \textit{Gnome GTK} project.
Our results show that \toolname~significantly outperforms both algorithms, achieving an $F_1$ score of 0.985.
\end{abstract}

\section{Introduction}
The ease of creating user accounts on the modern internet brings the peculiarity that a single individual can be registered under multiple different user accounts.
Conversely, it happens that two user accounts of entirely different individuals appear surprisingly similar, e.g. when the individuals have a common name.
Such ambiguities may occur for user accounts in a wide range of online services, be it social media, scientific publication databases, or version control systems for code.
If these ambiguities are not adequately resolved, subsequent studies on citation biases \citep{PhysRevE.102.032303, Zingg2020} or social relations \citep{Gote2019git2net, Sarigol2017, Pfitzner2014} can be significantly biased \citep{newman2004best}.
Therefore, various algorithms have been developed to disambiguate such user accounts by mapping an individual's different aliases to a unique identifier.
By ensuring that characteristics describing a single individual are not determined for the different aliases separately, name disambiguation algorithms represent integral tools in numerous scientific disciplines analysing real-world user data.

In this work, we consider communities of software developers who commit code to \textit{Version Control Systems} such as \textit{git}.
The author of a commit specifies their alias in the form of a user name and an email address.
Thereby, an author often commits code under different aliases.
Further, different authors can have almost identical aliases.
A variety of algorithms have been suggested that allow us to disambiguate such cases.
However, most algorithms require manually curated training data, which is costly to obtain for large sets of different repositories.
Further, the algorithms often rely on a wide range of features, including external sources and data about author behaviour, disqualifying this information for use in subsequent studies to avoid overfitting.
Finally, easily applicable reference implementations of the proposed algorithms are largely missing.

With the present work, we aim to pick up these points by making the following contributions:
\begin{itemize}
    \item We propose \toolname, an author disambiguation algorithm that uses only name and email data.
        Such a small amount of required information widens the number of cases to which author disambiguation can be applied.
        Thereby, other information can be reserved for subsequent analyses as it is not used already in a data preprocessing step.
    \item We evaluate \toolname~in a comparison study, showing that our method outperforms commonly used disambiguation algorithms by means of overall disambiguation accuracy. 
    \item We make \toolname~available as an easy-to-use Open Source Python package available via \textit{Pypi} (\texttt{pip install gambit-disambig}).
\end{itemize}

\section{Related Work}\label{sec:related_work}
Author disambiguation algorithms have been developed in multiple disciplines.
For example, in scientometrics, algorithms were developed to disambiguate author names in literature databases \citep{sinatra2016quantifying, kim2018evaluating}.
For empirical software engineering data, ranging from version control systems to mailing lists, a recent review of existing approaches can be found in \citep{wiese2016mailing}.

Disambiguation approaches can be divided into \textit{exogenous} and \textit{endogenous} approaches \citep{kouters2012s}.
\textit{Exogenous} approaches aim to optimise their predictive performance by collecting and analysing additional information not present in the analysed repository, e.g. GPG key, mailing lists, or maintained contributor lists \citep{robles2005developer}.
\textit{Endogenous} approaches, on the other hand, aim to perform the disambiguation task using only the information present in the repository.
The set of \textit{endogenous} methods can be further subdivided into \textit{learning-based} and \textit{learning free} approaches.
\textit{Learning-based} approaches have been shown to outperform \textit{learning free} approaches in a variety of cases \citep{ventura2015seeing, sarawagi2002interactive, amreen2020alfaa}.
However, a significant downside is that they require manually disambiguated training data for each repository, which is highly labour intensive to obtain.
Further, they often require additional information aside from names and email addresses to perform the disambiguation \citep{amreen2020alfaa}.

With \toolname, we propose a tool for author disambiguation in large sets of repositories for subsequent analysis of author- and team-behaviour.
This task requires methods that neither involve (i) prior training on manually disambiguated data nor (ii) information other than name and email.
Two rule-based methods fulfilling these criteria have been proposed by \citep{goeminne2013comparison} and \citep{bird2006mining}.
Both algorithms first perform an initial cleaning and preprocessing step and then extract the email base $EB$ consisting of the email $E$ up to the ``@'' symbol.
The ``Simple algorithm'' \citep{goeminne2013comparison} then matches two aliases if either their name $N$ or their email base $EB$ are an exact match.
Bird et al. \citep{bird2006mining} further derive the first $FN$ and last names $LN$ of aliases as the first and last part of the name $N$.
Using this information, a match between two aliases, $i$ and $j$, is detected if at least one of the following conditions holds:
\begin{itemize}
    \item $\text{\rmfamily sim}(N_i, N_j) \geq t$,
    \item $\min(\text{\rmfamily sim}(FN_i, FN_j),\text{\rmfamily sim}(LN_i, LN_j)) \geq t$,
    \item both $FN_{i(j)}$ and $LN_{i(j)}$ are in $EB_{j(i)}$,
    \item $FN_{i(j)}[0] + LN_{i(j)}$ is in $EB_{j(i)}$,
    \item $FN_{i(j)} + LN_{i(j)}[0]$ is in $EB_{j(i)}$,
    \item $\text{\rmfamily sim}(EB_i, EB_j) \geq t$.
\end{itemize}
Here, $\text{\rmfamily sim}$ refers to the Levenshtein similarity shown in \cref{eq:levenshtein_sim} and $t$ is an arbitrary threshold above which a match is detected.
$FN_{i(j)}[0]$ and $LN_{i(j)}[0]$ are the initial letters of the respective name, $+$ concatenates two strings, and $i(j)$ means that $i$ and $j$ are used interchangeably. 

While notably, the algorithm by Bird et al. \citep{bird2006mining} shows excellent baseline performance, it tends to detect false positives when email bases consist of common first names \citep{kouters2012s}.
This tendency has also been reported in the original paper, arguing that ``it is much easier during a manual step to split clusters than to unify two disparate clusters from a very large set'' \citep{bird2006mining}.

\section{gambit: Overview}\label{sec:gambit}
\begin{figure*}
        \centering
    \scalebox{.85}{
    \begin{tikzpicture}[node distance=.5cm]\sffamily\footnotesize
    	
    	\draw[ultra thick, fill=customG!20, draw=none, rounded corners=10] (5.8,.5) rectangle (13.2,-6.15);
    	\draw[ultra thick, fill=black!10, draw=none, rounded corners=10] (-1.4,.5) rectangle (5,-6.15);
    	\draw[ultra thick, fill=black!10, draw=none, rounded corners=10] (14,.5) rectangle (17.3,-6.15);
    	
    	\node[anchor=south] at (1.8,.5) {\textbf{Author} $i$\vphantom{$j$}};
    	\node[anchor=south] at (15.35,.5) {\textbf{Author} $j$};
    	\node[anchor=south] at (9.5,.5) {\textbf{Match if average of top two is greater or equal to $t$}\vphantom{$j$}};
    	
    	\draw[line width=2pt, white] (5.8, -0.6) -- (13.2, -0.6);
    	\draw[line width=2pt, white] (5.8, -3.37) -- (13.2, -3.37);
    	\draw[line width=2pt, white] (5.8, -5) -- (13.2, -5);

    	
    	\node[] (rec_name1) {\textbf{recorded name}};
    	\node[node distance=5.6cm, below=of rec_name1.center, anchor=center] (rec_email1) {\textbf{recorded email}};
    	
    	\node[node distance=5cm, right=of rec_name1.center, anchor=east] (name1) {\textbf{name} $N_i$};
    	\node[node distance=5cm, right=of rec_email1.center, anchor=east] (email1) {\textbf{email} $E_i$};
    	
    	\node[node distance=1.1cm, below=of name1.east, anchor=east] (first_name1) {\textbf{first name} $FN_i$};
    	\node[below=of first_name1.east, anchor=east] (middle_name1) {\textbf{penultimate name} $PN_i$};
    	\node[below=of middle_name1.east, anchor=east] (last_name1) {\textbf{last name} $LN_i$};
    	
    	\node[node distance=.8cm, above=of email1.east, anchor=east] (email_base1) {\textbf{email base} $EB_i$};
    	
    	
    	\node[node distance=9cm, right=of name1.east, anchor=west] (name2) {\textbf{name} $N_j$};
    	\node[node distance=9cm, right=of email1.east, anchor=west] (email2) {\textbf{email} $E_j$};
    	
    	\node[node distance=1.1cm, below=of name2.west, anchor=west] (first_name2) {\textbf{first name} $FN_j$};
    	\node[below=of first_name2.west, anchor=west] (middle_name2) {\textbf{penultimate name} $PN_j$};
    	\node[below=of middle_name2.west, anchor=west] (last_name2) {\textbf{last name} $LN_j$};

    	\node[node distance=.8cm, above=of email2.west, anchor=west] (email_base2) {\textbf{email base} $EB_j$};
    	
    	
    	\node[node distance=5cm, right=of name1.east, anchor=center,yshift=2mm] (sim_N) {$\text{\rmfamily sim}(N_i, N_j)$};
    	\node[node distance=0cm, below=of sim_N.south, anchor=north] (eq_N) {$N_i$ is equal to $N_j$};
    	\node[node distance=0cm, below=of eq_N.south, anchor=north, align=center] (sim_FF) {$\min(\text{\rmfamily sim}(FN_i, FN_j),\max(\text{\rmfamily sim}(LN_i, LN_j),$\\$ \text{\rmfamily sim}(LN_i, PN_j), \text{\rmfamily sim}(PN_i, LN_j)))$};
    	\node[node distance=0cm, below=of sim_FF.south, anchor=north, align=center] (sim_FL) {$\min(\text{\rmfamily sim}(FN_i, LN_j),\max(\text{\rmfamily sim}(PN_i, FN_j,$\\$ \text{\rmfamily sim}(LN_i, PN_j), \text{\rmfamily sim}(LN_i, FN_j)))$};
    	\node[node distance=0cm, below=of sim_FL.south, anchor=north, align=center] (sim_LF) {$\min(\text{\rmfamily sim}(LN_i, FN_j),\max(\text{\rmfamily sim}(PN_i, LN_j,$\\$ \text{\rmfamily sim}(FN_i, PN_j), \text{\rmfamily sim}(FN_i, LN_j)))$};
    	
        \node[node distance=0cm, below=of sim_LF.south, anchor=north, align=center] (in_PF1) {$FN_{i(j)}[0] + LN_{i(j)}$ is in $EB_{j(i)}$};
        
        \node[node distance=0cm, below=of in_PF1.south, anchor=north, align=center] (in_PL1) {$FN_{i(j)} + LN_{i(j)}[0]$ is in $EB_{j(i)}$};
        
        \node[node distance=0cm, below=of in_PL1.south, anchor=north, align=center] (in_FL1) {$2 \times$ \{both $FN_{i(j)}$ and $LN_{i(j)}$ are in $EB_{j(i)}$\}};

    	\node[node distance=0cm, below=of in_FL1.south, anchor=north, align=center] (eq_E) {$2 \times$ \{$E_i$ is equal to $E_j$\}};
    	\node[node distance=0cm, below=of eq_E.south, anchor=north, align=center] (sim_EB) {$\text{\rmfamily sim}(EB_i, EB_j)$};
    	
    	\node[anchor=east] at (6.7cm, 0 |- sim_N) {\textbf{1.}};
    	\node[anchor=east] at (6.7cm, 0 |- eq_N) {\textbf{2.}};
        \node[anchor=east] at (6.7cm, 0 |- sim_FF) {\textbf{3.}};
        \node[anchor=east] at (6.7cm, 0 |- sim_FL) {\textbf{4.}};
        \node[anchor=east] at (6.7cm, 0 |- sim_LF) {\textbf{5.}};
        \node[anchor=east] at (6.7cm, 0 |- in_PF1) {\textbf{6.}};
        \node[anchor=east] at (6.7cm, 0 |- in_PL1) {\textbf{7.}};
        \node[anchor=east] at (6.7cm, 0 |- in_FL1) {\textbf{8.}};
        \node[anchor=east] at (6.7cm, 0 |- eq_E) {\textbf{9.}};
        \node[anchor=east] at (6.7cm, 0 |- sim_EB) {\textbf{10.}};

    	
    	\draw[-latex] (rec_name1) -- node[midway, above, font=\footnotesize\itshape] {preprocessing} (name1);
    	
    	\draw[-latex] (rec_email1) -- node[midway, below, font=\footnotesize\itshape] {preprocessing} (email1);
    	
    	\draw[-latex, rounded corners] ($(name1.south east) - (1,0)$) --++ (0,-.15) --++ (-2.7,0) |- (first_name1.west);
    	\draw[-latex, rounded corners] ($(name1.south east) - (1,0)$) --++ (0,-.15) --++ (-2.7,0) |- (middle_name1.west);
    	\draw[-latex, rounded corners] ($(name1.south east) - (3.7,.35)$) |- node[pos=.25, left, font=\footnotesize\itshape, yshift=1, align=right] {first,\\ second to last,\\  and last element\\ after split at\\ whitespaces} (last_name1.west);
    	
    	\draw[-latex, rounded corners] ($(email1.north east) - (1,0)$) --++ (0,.15) --++ (-2,0) |- node[pos=.25, left, font=\footnotesize\itshape] {email before ``@''} (email_base1.west);
    	
    	
    	\draw[-latex] (name1.east) --++ (.8,0);
    	\draw[-latex, rounded corners] (first_name1.east) --++ (.8,0);
    	\draw[-latex, rounded corners] (middle_name1.east) --++ (.8,0);
    	\draw[-latex, rounded corners] (last_name1.east) --++ (.8,0);
    	\draw[-latex, rounded corners] (email_base1.east) --++ (.8,0);
    	\draw[-latex, rounded corners] (email1.east) --++ (.8,0);
    	
    	\draw[-latex, rounded corners] (first_name1.east) --++ (.5,0) --++ (0,-2.5) --++ (.3,0);
    	\draw[-latex, rounded corners] (last_name1.east) --++ (.3,0) --++ (0,-1.8) --++ (.5,0);
    	\draw[-latex, rounded corners] (email_base1.east) --++ (.4,0) --++ (0,-.47) --++ (.4,0);
    	
    	\draw[-latex] (name2.west) --++ (-.8,0);
    	\draw[-latex, rounded corners] (first_name2.west) --++ (-.8,0);
    	\draw[-latex, rounded corners] (middle_name2.west) --++ (-.8,0);
    	\draw[-latex, rounded corners] (last_name2.west) --++ (-.8,0);
    	\draw[-latex, rounded corners] (email_base2.west) --++ (-.8,0);
    	\draw[-latex, rounded corners] (email2.west) --++ (-.8,0);
    	
    	\draw[-latex, rounded corners] (first_name2.west) --++ (-.5,0) --++ (0,-2.5) --++ (-.3,0);
    	\draw[-latex, rounded corners] (last_name2.west) --++ (-.3,0) --++ (0,-1.8) --++ (-.5,0);
    	\draw[-latex, rounded corners] (email_base2.west) --++ (-.4,0) --++ (0,-.47) --++ (-.4,0);
    	
    \end{tikzpicture}
    }
    \caption{Overview of \toolname. After a preprocessing step, features are extracted from name and email information, and similarities are computed based on ten rules. Two authors are matched if the average of the top two similarities matches or exceeds a threshold $t$.}
    \label{fig:algorithm_overview}
\end{figure*}
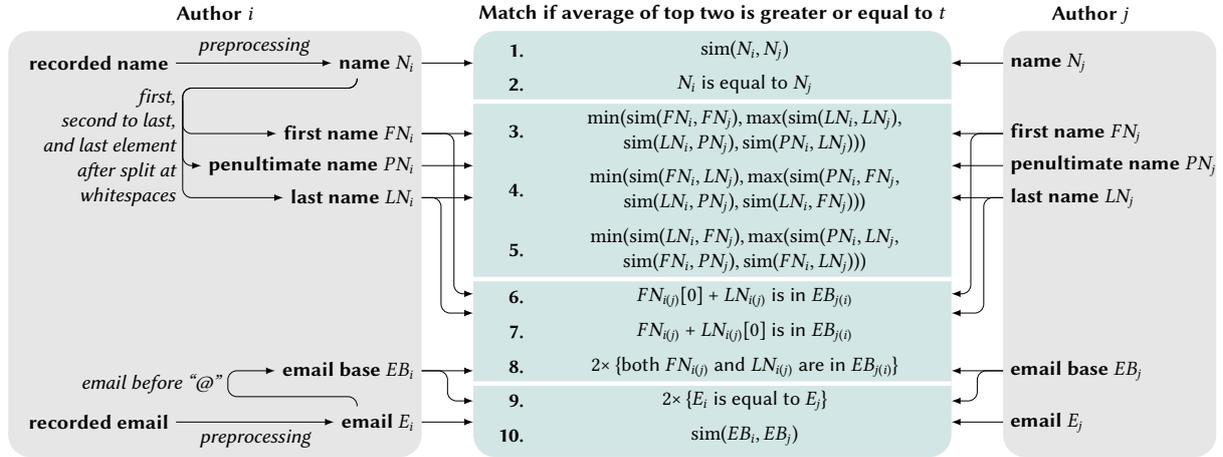
With \toolname, we propose a disambiguation algorithm that (i) requires no training, (ii) relies only on name and email data, while (iii) simultaneously reducing the number of false positives compared to existing algorithms.
\toolname~operates in four consecutive steps that we illustrate in \cref{fig:algorithm_overview} and describe in the following.
Please note that our algorithm has been developed based on the \textit{Apache Hadoop} repository\footnote{\url{https://github.com/apache/hadoop}}, but its empirical evaluation is performed for the \textit{Gnome GTK} project\footnote{\url{https://github.com/gnome/gtk}} to avoid the potential of overfitting in the results.
\paragraph{Preprocessing}
In an initial preprocessing step, both the name and email are cleaned from characters or strings that impede the subsequent matching process.
Non-ASCII characters are mapped to their closest ASCII counterpart.
Delimiters such as ``+'', ``-'', ``,'', ``.'', ``\_'', ``;'', as well as all instances of camel case are replaced with whitespaces.
All remaining non-alphabetical characters are removed, except whitespaces and the ``@'' symbol.
All text is converted to lower case, and names of time zones and common strings such as ``jr'' or ``admin'' are removed to obtain an entity's name $N$ and email $E$ shown in \cref{fig:algorithm_overview}.
\paragraph{Entity Extraction}
We then extract additional entity matching features from both the name and email information.
These include first name $FN$, penultimate name $PN$, and the last name $LN$ by selecting corresponding whitespace-separated elements of the name $N$.
We extract the email base as all characters in $E$ leading up to the ``@'' sign.
\paragraph{Similarity Computation}
We determine the similarity between different entities based on a set of ten rules.
To compare two strings, $s_1$ and $s_2$, existing algorithms commonly use the normalised Levenshtein edit distance \citep{bird2006mining, kouters2012s}
\begin{align}
    \text{sim}_{lev}(s_1, s_2) = 1-\frac{d_{lev}(s_1, s_2)}{\max(|s_1|, |s_2|)}.\label{eq:levenshtein_sim}
\end{align}
Here, $d_{lev}(s_1, s_2)$ denotes the Levenshtein edit distance \citep{levenshtein1966binary} between the two strings, and $|s|$ counts the number of characters.
The authors of \cite{amreen2020alfaa} found that, for their algorithm, the normalised Levenshtein edit distance was outperformed by the Jaro-Winkler similarity defined as
\begin{align*}
    \text{sim}_{jw}(s_1, s_2) &= \text{sim}_j(s_1, s_2) + 0.1 l(1-\text{sim}_j(s_1, s_2)),\\
    \text{sim}_j(s_1, s_2) &= \begin{cases} 0 & \text{for } c=0,\\ \frac{1}{3}\left(\frac{c}{|s_1|} + \frac{c}{|s_2|} + \frac{c-\tau}{c}\right) & \text{for } c\neq 0.\end{cases}
\end{align*}
Here, $c$ is the number of common characters between the strings, $\tau$ is the number of character transpositions, and $l$ is the length of a common prefix in $s_1$ and $s_2$ up to at most four characters \citep{winkler1990string}.
We will compare both measures when evaluating our method in \cref{sec:quality_performance}.

The rules based on which \toolname~matches identities shown in \cref{fig:algorithm_overview} extend the rules proposed by \citep{bird2006mining}.
Rules 1-2 compare the full names in terms of similarity as well as equality.
With rules 3-5, we compare first, last, and penultimate names.
Following the suggestion by \citep{amreen2020alfaa}, we also account for the potential inversion of names.
Rules 6-8 compare the name of an alias $i$ to the email base of $j$.
We match $i$ and $j$ under three conditions:
(i) if the initial of $FN_i$ prepended to $LN_i$ is a substing of $EB_j$,
(ii) if $FN_i$ prepended to the initial of $LN_i$ is a substing of $EB_j$,
(iii) if both $FN_i$ and $LN_i$ are present in $EB_j$.
Finally, rules 9-10 compare both the full email as well as the email base.
Similarities are only computed if both compared strings are at least three characters long to prevent coincidental matches, which are more likely for very short strings.
If at least one string is shorter, we directly consider the two strings to be different.
Boolean matches are considered to be 1 when true and 0 when false.

\paragraph{Alias Matching}
The result of the similarity computation is a list of similarity values for a given pair of aliases.
Previous approaches commonly match two aliases if at least one of these similarities surpasses a given threshold $t$ \cite{bird2006mining, goeminne2013comparison}.
However, this is prone to detect \textit{false positives} with commonly used names.
To improve the robustness of our approach, \toolname~requires the average of the top two similarities to exceed a similarity threshold $t$.
Nevertheless, there are three cases in which a match should be detected even if only a single similarity surpasses $t$:
(i) in case the full names are identical, (ii) in case the full emails are identical, and (iii) if both the first name and the last name of an alias appear in the email base of another alias.
Case (i) is ensured by accounting for both name similarity and name equality in rules 1-2.
For cases (ii) and (iii), the similarity computed by the corresponding rules 8 and 9 is multiplied by a factor of 2.

When matching all identities in a repository, we further make use of the transitive property of similarity.
Hence, if alias $A$ is matched with aliases $B$ and $C$, we directly consider $B$ and $C$ to match as well.

\section{Quality and Performance}\label{sec:quality_performance}
\setlength{\w}{5.05cm}
\setlength{\h}{4.6cm}

\begin{figure*}
    \centering
    \scalebox{.85}{
    \input{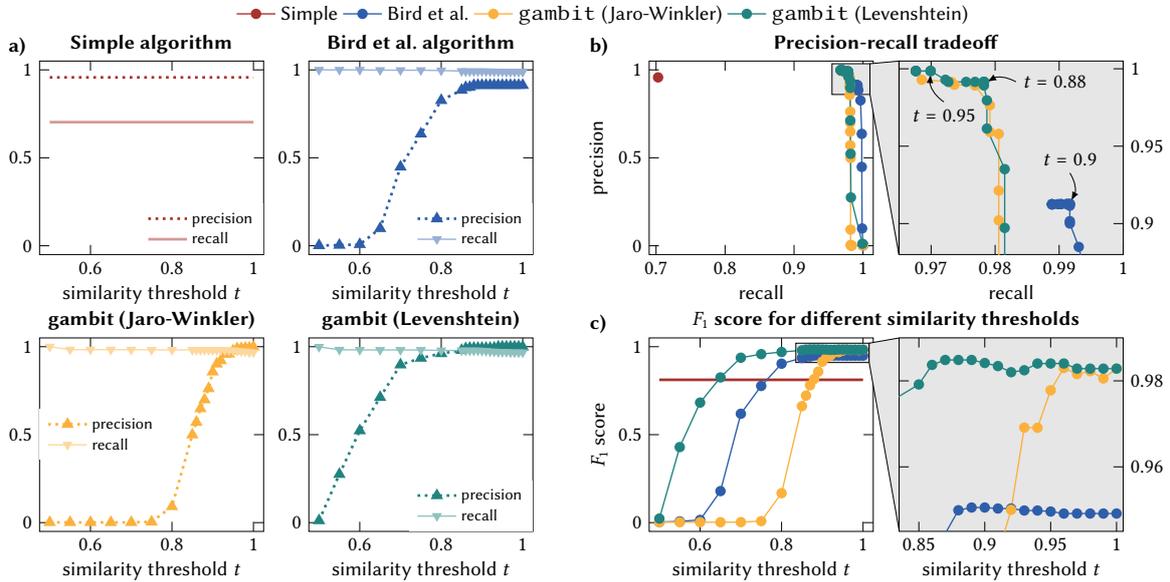}
    }
    \caption{Disambiguation performance for the \textit{Gnome GTK} project against a manually disambiguated baseline. {\sffamily\textbf{a)}} shows precision and recall for different similarity thresholds $t$. {\sffamily\textbf{b)}} presents the tradeoff between precision and recall, and {\sffamily\textbf{c)}} compares all algorithms based on the $F_1$ score.}
    \label{fig:disambiguation_results}
\end{figure*}
We compare \toolname~to the two other commonly used rule-based disambiguation approaches that only rely on name and email information: the Simple algorithm \citep{goeminne2013comparison} and the method proposed by Bird et al. \citep{bird2006mining}.
Following past studies \citep{kouters2012s, goeminne2013comparison}, we used the \textit{Gnome GTK} project\footnote{\url{https://github.com/gnome/gtk}, cloned on 2020-10-09 14:22 GMT} to evaluate and compare the different disambiguation techniques.
As a first step, we created a disambiguated ground-truth data set.
Creating such a data set entirely manually for \textit{Gnome GTK} is infeasible, as the repository contains 1896 unique name-email pairs, requiring 1.8 million comparisons.
We, therefore, opted for a semi-automated approach that automatically matches identities with identical names or email addresses.
We further automatically marked all identities with a normalised Levenshtein distance below 0.5 for both name and email as different.
The remaining 10074 identities were manually disambiguated independently by the two authors of the present manuscript over multiple days.
In total, we achieved a close-to-perfect inter-rater agreement \citep{cohen1960coefficient, landis1977application} of $\kappa = 0.994$ for this manual subset.
Despite this, the automated part can still lead to wrong classifications, particularly for short strings where the normalised Levenshtein distance shows higher variability with changes of individual characters.
Therefore, we ensured transitivity for all matches and manually checked all automated matches for strings with less than six characters.
Furthermore, we manually confirmed a subset of matches for which errors with the tested algorithms were observed.
Overall, this resulted in a ground-truth data set without any obvious errors, containing 2161 ambiguities.

Subsequently, we applied all three algorithms to the name-email pairs and evaluated the results against the manual ground-truth.
For \toolname, we evaluated the Jaro-Winkler similarity and the normalised Levenshtein distance as similarity measures.
The simple algorithm does not have any hyper-parameters.
For both \toolname and the algorithm proposed by Bird et al., we tested multiple similarity thresholds $t$ between 0.5 and 1.
The computation times for each threshold were approximately 4, 7, and 12 minutes for the Simple, Bird et al., and \toolname~algorithms, respectively\footnote{Intel\textregistered~Core\texttrademark~i9 7960X, 16C/32T, 2.80GHz base, 4.2GHz boost}.

\Cref{fig:disambiguation_results} shows the results.
In \cref{fig:disambiguation_results}a, values for precision $p = \nicefrac{T_p}{T_p+F_p}$ and recall $r=\nicefrac{T_p}{T_p+F_n}$, computed using \textit{scikit-learn} \citep{scikit-learn}, are shown for different similarity thresholds $t$.
Here, $T_p$, $F_p$, and $F_n$ represent the number of true positive, false positive, and false negative classifications compared to the ground-truth.
An ideal algorithm simultaneously maximises precision and recall.
The tradeoff for our algorithms is depicted in \cref{fig:disambiguation_results}b.
We can see that the Simple algorithm is significantly worse than both \toolname~and Bird et al.'s algorithm in terms of recall.
Further, while the algorithm described by Bird et al. slightly outperforms \toolname~in terms of recall, \toolname~detects significantly fewer $FP$s with only a few more $FN$s.
In other words, \toolname~increases precision with only slight decreases in recall.
This can also be seen in the $F_1~\text{score} = \nicefrac{2pr}{p+r}$, which is largest for \toolname.
This score is the harmonic mean between precision $p$ and recall $r$, and is shown in \cref{fig:disambiguation_results}c.

When comparing \toolname~using Jaro-Winkler similarity and normalised Levenshtein distance, we find that the normalised Levenshtein distance reaches high $F_1$ scores for a broader range of similarity thresholds $t$.  
This is due to the Jaro-Winkler similarity primarily considering the similarity between the start of strings, whereas the Levenshtein distance considers the entire strings.
This focus leads to generally higher similarity scores for Jaro-Winkler, thus requiring a higher similarity threshold $t$
In terms of $F_1$ score, similarity thresholds above 0.85 all yield excellent results.
We expect this finding to generalise to other repositories than \textit{Gnome GTK}; however, we were not able to confirm this due to the large amount of manual labour involved with generating additional ground-truth data sets.
Overall, we recommend using \toolname~with the normalised Levenshtein distance and a similarity threshold of $t=0.95$.

\section{Threats to validity}\label{sec:threats_to_validity}
Creating a manual ground-truth baseline is a highly labour intensive and error-prone process.
Furthermore, the final decision to match two aliases, e.g. based on identical names, is subjective.
Two aliases with identical names do not necessarily disambiguate to the same author, and authors use different names in different contexts or stages of their lives \citep{svajlenko2016machine, wiese2016mailing}.
Therefore, it is impossible to guarantee that all entries in our ground-truth data are correct.
We have made all efforts to overcome these issues and make our data available for replication studies; however, the manual ground-truth baseline represents the biggest threat to validity for our results.
Unfortunately, this limitation could only be fully resolved with additional data on the true aliases that, to the best of our knowledge, does not exist.
Therefore, using a manual baseline represents the best available approach for our study.

Further, due to the large amounts of manual labour involved with creating ground-truth data, we have only performed our evaluation for a single project.
Therefore, at this stage, we cannot make any performance claims for other projects.
However, we note that the evaluation was performed on data not used in the creation of \toolname, hence removing the potential of overfitting for this data set.

\section{Conclusion and Outlook}\label{sec:conclusion}
Name disambiguation is a complex but highly relevant challenge whenever analysing real-world user data, such as data from version control systems.
With \toolname, we propose a rule-based disambiguation algorithm that only relies on name and email information, hence allowing for a wide area of applications.
We carefully evaluate its performance against the ``Simple algorithm'' \citep{goeminne2013comparison} and an algorithm proposed by Bird et al. \citep{bird2006mining}, two commonly used algorithms with similar characteristics on manually disambiguated ground-truth data from the \textit{Gnome GTK} project.
Our results show that \toolname~significantly outperforms both algorithms in terms of precision and $F_1$ score.
Not relying on external information or manually curated training sets makes our algorithm highly scalable and enables analyses of large sets of OSS projects.
\toolname~is publicly available and easily accessible as a Python package.

In future work, we will compare \toolname's performance against exogenous and learning-based algorithms and extend this study to a more extensive set of projects.

\section*{Tool Availability, Archival, and Reproducibility}
\toolname~is available via \textit{PyPI} (\texttt{pip install gambit-disambig}) and as OSS project on GitHub\footnote{\url{https://github.com/gotec/gambit}}.
To facilitate the disambiguation of \textit{git} repositories, \toolname~is also integrated in \texttt{git2net}, an Open Source Python package to mine co-editing networks from \textit{git} repositories \citep{Gote2019git2net}.

Our implementations of all algorithms used in the evaluation are archived on \url{zenodo.org}\footnote{\zenodolink}. Please contact us directly for access to the manually disambiguated ground-truth data.

\bibliographystyle{sg-bibstyle}
\bibliography{bibliography}

\end{document}